\documentclass{article}
\usepackage{amsmath,graphicx}
\usepackage{gensymb}
\usepackage{authblk}

\title{Group-wise 3D registration based templates to study the evolution of ant worker neuroanatomy}
\author[1,2,3]{Ignacio Arganda-Carreras}
\author[4]{Darcy G Gordon}
\author[4,5]{Sara Arganda}
\author[6]{Maxime Beaudoin}
\author[4]{James FA Traniello}
\affil[1]{Ikerbasque, Basque Foundation for Science, Bilbao, Spain}
\affil[2]{Dept. of Computer Science and Artificial Intelligence, Basque Country University, San Sebastian, Spain}
\affil[3]{Donostia International Physics Center (DIPC), San Sebastian, Spain}
\affil[4]{Department of Biology, Boston University, Boston, Massachusetts, USA}
\affil[5]{Centre de Recherches sur la Cognition Animale (CRCA), CNRS, Toulouse, France}
\affil[6]{Ecole d'Ingenieurs et Centre de Recherche (ENSTA), Bretagne, France}

\date{ }

\begin{document}

\maketitle
\begin{abstract}
The evolutionary success of ants and other social insects is considered to be intrinsically linked to division of labor and cooperative behavior, including task specialization and emergent collective intelligence.  Selection for both individual- and colony-level behavioral performance concerns the role of the brains of individual workers in generating behavior, but how the ``social brain'' is structured is poorly understood. Confocal imaging and manual annotations of brain scans have been used to understand the mosaic organization of the ant brain by quantifying brain regions volumes. These studies require laborious effort and may be subject to potential bias. To address these issues and increase throughput necessary to robustly sample ant species diversity and to perform evolutionary analyses, we propose a group-wise 3D registration approach to build for the first time bias-free brain atlases of intra- and inter-subcaste individuals and automatize the segmentation of new individuals.  Generating brain templates will accelerate data collection and greatly expand research opportunities in the study of the evolutionary neurobiology of ants and other social insects.

\end{abstract}
%
%\begin{keywords}
%Image registration, image segmentation, neuroanatomy, template construction, social insects
%\end{keywords}
%
\section{Introduction}
\label{sec:intro}
Ants (\textit{Hymenoptera: Formicidae}), renowned for their remarkable diversity and ecological significance \cite{wilson1999diversity}, typically display extraordinary collective behavior \cite{holldobler2009superorganism}.  A key question in evolutionary neurobiology concerns how ant sociality, ecology, and the ability to make accurate group decisions have impacted their brain structure
The emergence of eusociality and social complexity are major novelties likely involving rapid behavioral changes that might be reflected in the anatomy of the brain \cite{ott2010gregarious,amador2015specialization}, although this idea has been controversial \cite{farris2013evolution,farris2016insect}. The remarkable evolutionary and ecological success of ants is hypothesized to be due to their social organization, which features division of labor, and collective behavior \cite{wilson1987causes}.

Workers in ant colonies are so intrinsically interdependent that they are considered superorganisms. The ``brain'' of such a superorganism evolved at two levels: to enable individual workers to respond adaptively as individuals acting independently of other workers, and colonies behaving as decision-making groups to cope with the multiple challenges of sociality (coordinated foraging, task specialization, communication, social interactions, nestmate recognition, e.g.). The Social Brain Hypothesis, originally postulated for primates, posits that individual members of larger groups require bigger brains to adaptively process social information \cite{dunbar2007evolution}. However, the degree to which this hypothesis can be meaningfully applied to eusocial insects has been debated  \cite{lihoreau2012exploration}. Brain evolution in ants, for example, must have evolved in consideration of body size, and therefore miniaturization of the nervous system.In addition to that, collective intelligence and division of labor may have relaxed individual cognitive challenges \cite{gronenberg2009social}. However, it is unclear if social selection favored the evolution of allometrically smaller or larger brains, as both patterns have been described \cite{riveros2012evolution,kamhi2016social}

The ant brain is a mosaic of different subregions (neuropils) that serve different functions  \cite{strausfeld2012atlas}: sensory perception (antennal and optic lobes), motor control and navigation (central body and subesophageal ganglion), and multi-sensorial integration, learning and memory (mushroom bodies). Using confocal imaging and manual annotations of brain regions, Muscedere \textit{et al.} demonstrated that minor and major workers of different ages of three species of \textit{Pheidole} have distinct patterns of brain size variation \cite{muscedere2012division}. These differences in subregion sizes and scales reflect the intra-colony division of labor and the sociobiological characteristics of this species. However, all these results come at the cost of allocating significant time to manual record the volumes of functionally specialized brain compartments, which may introduce a bias.

Recent advances in image processing, inspired in techniques developed to study the human brain, have allowed extraordinary outputs of unprecedented quality and throughput in neuroanatomical studies in honeybees \cite{rybak2012digital} and fruit flies \cite{rein2002drosophila,Costa2016} among other insects \cite{Menzel2012}. These approaches combine multiple brains in a single model or template, which statistically represents the whole species. Replication is necessary to avoid biases originated in the fixation and imaging processes of the brains as well as to account for inter-individual variability. 

Template brains have a dual function. Transforming all samples to the same reference space allows normalizing the information from brains imaged under different conditions or image modalities, and anatomical regions of reference brains are usually annotated, which produces the automatic segmentation of registered samples.Although many strategies have been proposed and evaluated in the last decades for the construction of brain templates in mammals \cite{talairach1988co, evans19933d, mazziotta1995probabilistic, chen2006neuroanatomical, dogdas2007digimouse, shattuck2008construction}, only a few of them have been applied to insect brains, most of them to \textit{Drosophila} data \cite{jefferis2007comprehensive, Yu2010, cachero2010sexual,Costa2016}. However, these results have not been translated yet into the ant brain community. This can be partially explained by the lack of expert-made anatomical labels and the larger morphological variability existing in the ant brain, what substantially hinders the registration process.

To address these issues, we propose a two-step co-registration solution that allows the construction of atlases of intra- and inter-caste individuals and identify specific differences between anatomical regions. Moreover, we have evaluated our approach in a total of $50$ labeled brains of four species of \textit{Pheidole}, a hyperdiverse genus of ants that exhibits striking morphological differentiation and division of labor: complete dimorphism or “trimorphism” in the worker caste. 

\section{Materials and Methods}
\label{sec:materials}

\subsection{Ant brains dataset}

\subsubsection{Ant species}
\textit{Pheidole}, the most diverse and species rich ant genus \cite{wilson1985ants}, is characterized by worker polymorphism (minor worker, major workers and, in some species, supersoldiers). Four \textit{Pheidole} species, courtesy of Dr. Diana Wheeler’s lab at the University of Arizona, have been selected for this study: \textit{P. spadonia}, \textit{P. rhea}, \textit{P. tepicana} and \textit{P. obtusospinosa}.
\subsubsection{Brain imaging and labeling}
\begin{figure}[htb]
\centering
\centerline{\includegraphics[width=12cm]{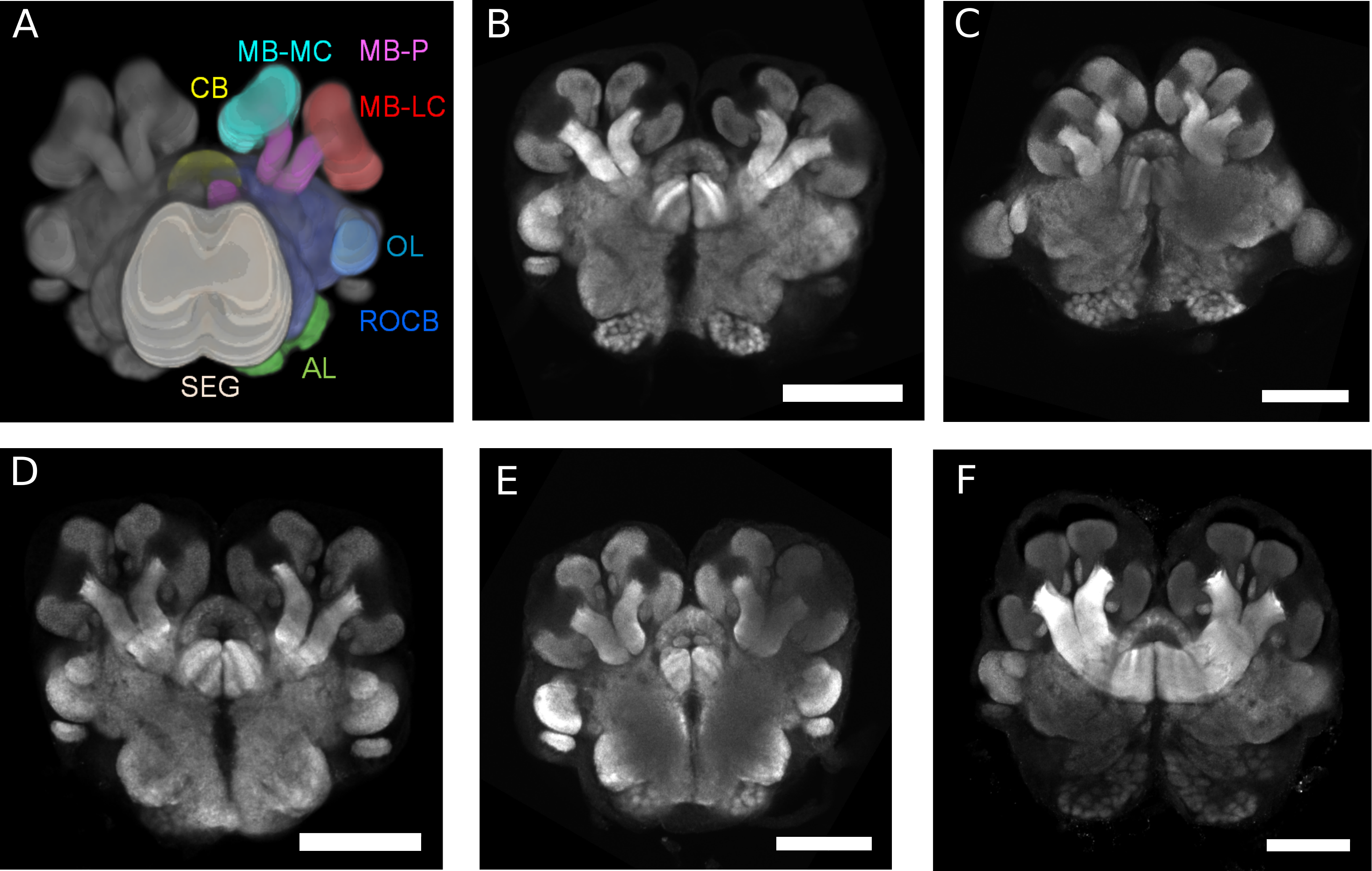}}
\caption{Examples of brain samples and labeled regions. From left to right and from top to bottom: 3D view of anatomical regions (A) and central sections of \textit{P. spadonia} minor (B),  \textit{P. spadonia} minor (C), \textit{P. tepicana} (D), \textit{P. obtusospinosa} (E) and \textit{P. rhea} (F) samples.  Scale bar: $100\mu m$.}
\label{fig:brains-and-labels}
\end{figure}

The immunohistochemical staining and imaging of ant brain neuropil was slightly modified from \cite{ott2008confocal, muscedere2012division}. We imaged 50 brains at a resolution of $\sim0.7\times0.7\times5 \mu$m/voxel: 10 minor worker brains from the mentioned four species and 10 major worker brains from \textit{P. spadonia}. Right brain hemispheres were manually labeled by an expert into $8$ anatomical regions: optic lobes (OL), antennal lobes (AL), mushroom body medial calyx (MB-MC), mushroom body lateral calyx (MB-LC), mushroom body peduncle (MB-P), central body (CB), subesophageal ganglion (SEG) and rest of the brain (ROCB). Fig.~\ref{fig:brains-and-labels} shows a 3D representation of the labels and brain samples of each type. (Image size: $\sim600\times600\times80$ pixels.)

\subsection{Image registration and template generation}
Group-wise templates were constructed using an algorithm building an average shaped brain within the diffeomorphic space. The approach uses symmetric diffeomorphic image registration (SyN) \cite{avants2008symmetric} with mutual information and cross-correlation to register a group of brain images to one another. The co-registration process is refined using a two-step strategy. First, all of the images are registered to one brain using only an affine transformation model and mutual information as the similarity measure to optimize. The resulting images are then averaged to form an initial blurry reference brain image. Second, the original brain images are non-linearly registered to this average to create a new average that maximizes the cross-correlation of the intensities of all brains. In this second step, the registration is improved gradually at different (in the present case, four) resolution levels and the result is an optimal average template.

For combining the co-registered images, we experimented first with a normalized voxel-wise average followed by sharpening with a Laplacian kernel (state-of-the-art in MRI). However, we found experimentally that an alternative strategy in which the template intensity image was generated by computing a voxel-wise median over the co-registered images produced slightly better results. The anatomical label image of the template was obtained by applying to each individual label image the diffeomorphic transformations computed from the corresponding confocal image, followed by a per-voxel majority voting over all warped label images.

Individual brain images were registered against the templates using the same two-step strategy, which performs an initial affine registration with mutual information as similarity metric followed by non-rigid registration with SyN and cross-correlation as similarity measure. The first registration is crucial in order to compensate for the large disparities in size among the different ant species and subcastes, while the second one locally finds an optimal solution.
All methods are implemented within the Advanced Normalization Tools (ANTs) software \cite{avants2011reproducible}.

\subsection{Evaluation metrics}

To evaluate the template performance, we registered test brains (not used in the template construction) against the template and transformed the template labels onto the test brain space.

We quantified the overlap ratio of the labels in the test brain space using the Dice similarity index, which provides a normalized measure of the overlap between two labels $L_i^A$ and $L_i^B$.
The Dice index is defined as
$$\mathrm{Dice}(L_i)=2\frac{|L_i^A\cap R_i^B|}{|L_i^A|+|L_i^B|}$$
where $|L_i^A|$ and $|L_i^A|$ are respectively the number of voxels of label $i$ in brain $A$ and $B$.

To quantify the shape and boundary errors, we measure the mean symmetric Euclidean distance between the surfaces of the labels. For each label $L_i$ in the pair of brains $A$ and $B$, we calculated the mean Euclidean distance $d_i^{A,B}$ between each surface point on $L_i^A$ and the closest surface point on $L_i^B$. The symmetric distance $d_i^{B,A}$ was calculated in an analogous way. The mean symmetric Euclidean distance was defined as
$$\mathrm{Mean\ Symmetric\ Euclidean\ distance}(L_i) = \frac{d_i^{A,B}+d_i^{B,A}}{2}$$

Finally, to measure the maximal boundary and shape differences between the original brain labels and the registered template labels, we calculated the mean symmetric Hausdorff distance. The Hausdorff distance $h_i^{A,B}$ of labels $L_i^A$ and $L_i^B$ is defined as the longest distance between any point on the surface of $L_i^A$ and the closest point on the surface of $L_i^B$. By computing $h_i^{B,A}$ in an analogous way, the symmetric Hausdorff distance can be calculated as 
$$\mathrm{Symmetric\ Hausdorff\ distance}(L_i) = \frac{h_i^{A,B}+h_i^{B,A}}{2}$$

Notice both distance metrics are expressed in absolute distance units.

\section{Results}
\subsection{Building an ant brain template}
\begin{figure}[htb]
\centering
\centerline{\includegraphics[width=6.5cm]{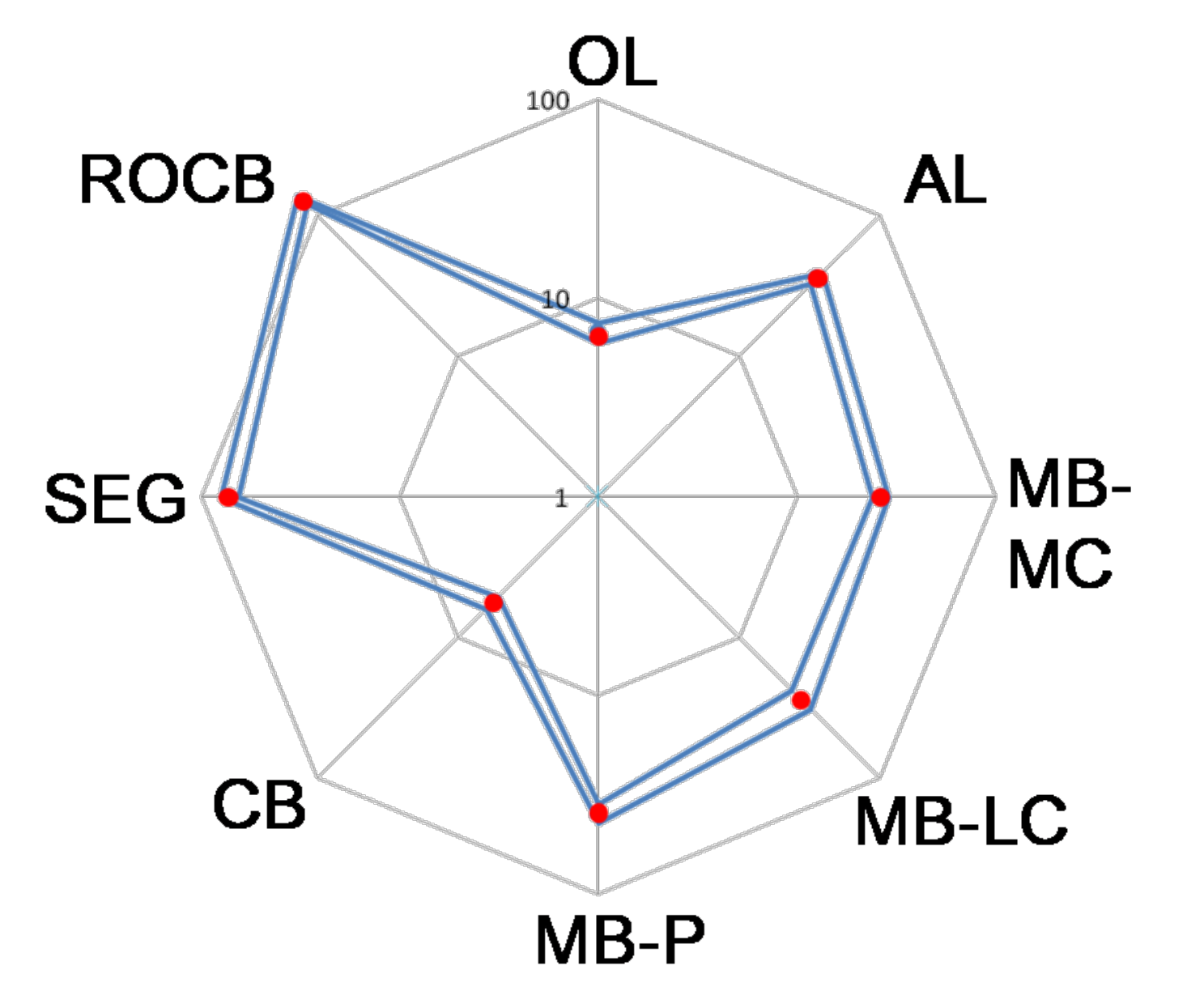}}
\caption{Evaluation of volume ($\times10^4 \mu m^3$) per anatomical region of minor \textit{P. spadonia} template. Blue lines represent the standard deviation of the volume of the original manual labels while the red dots are the template volume value.}
\label{fig:minor-volume}
\end{figure}

As a proof-of-concept of our methodology, we first attempted to build intra-species and intra-subcaste templates. For that reason, we chose the $10$ minor and $10$ major worker samples from \textit{P. spadonia}. Here we realized that an initial affine pre-registration was needed due to the volume variability and imaging conditions. Both templates were successfully built and evaluated based on how well their consensus labels represented the sample population (see Fig. \ref{fig:minor-volume}). 

\subsection{Building and evaluating hybrid templates}
\begin{figure}[htb]
\centering
\centerline{\includegraphics[width=12cm]{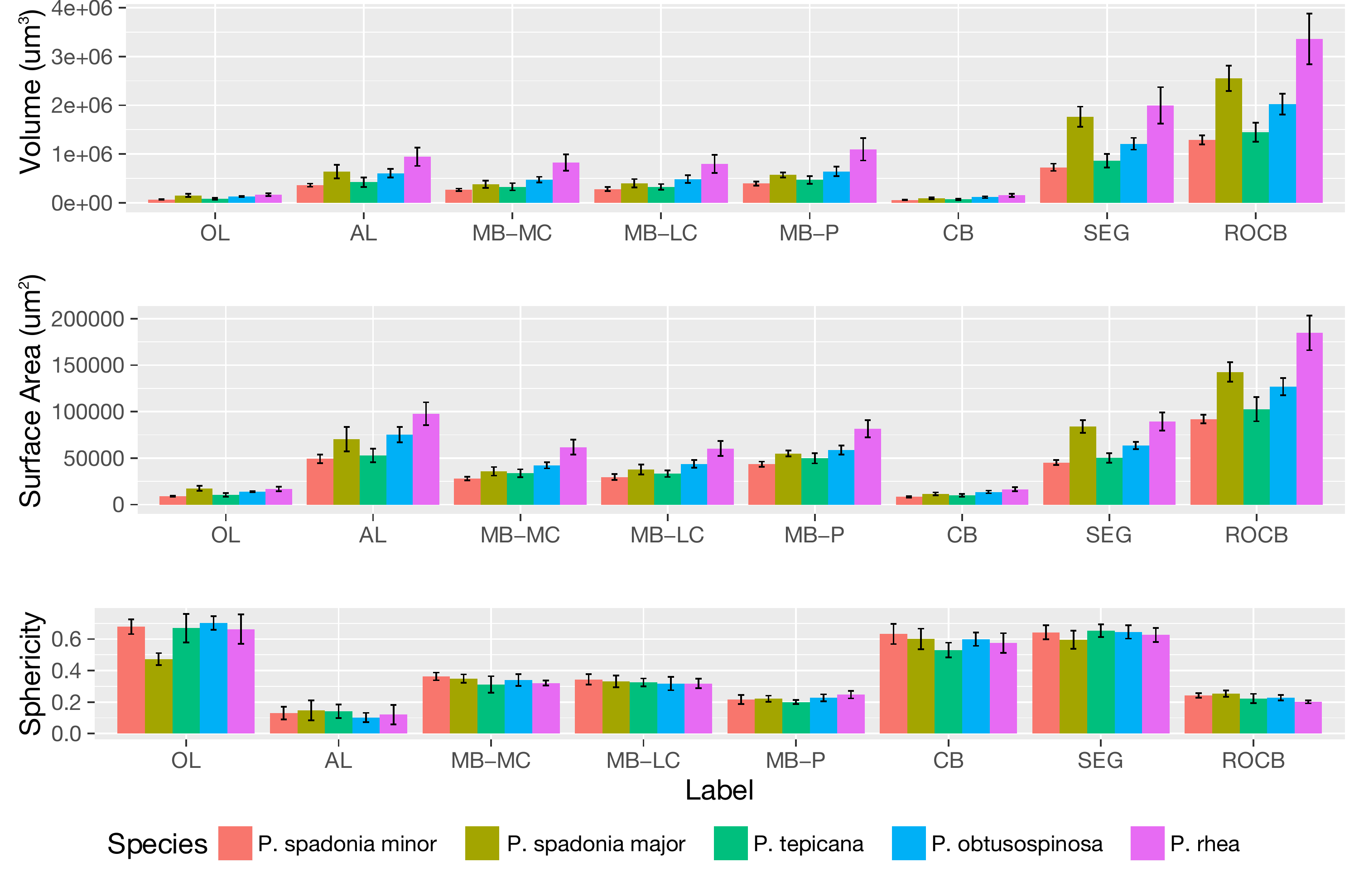}}
\caption{Morphological differences between species and subcastes. From top to bottom: volume, surface area and sphericity measurements.}
\label{fig:morphology-measures}
\end{figure}

\begin{figure}[htb]
\centering
\centerline{\includegraphics[width=12cm]{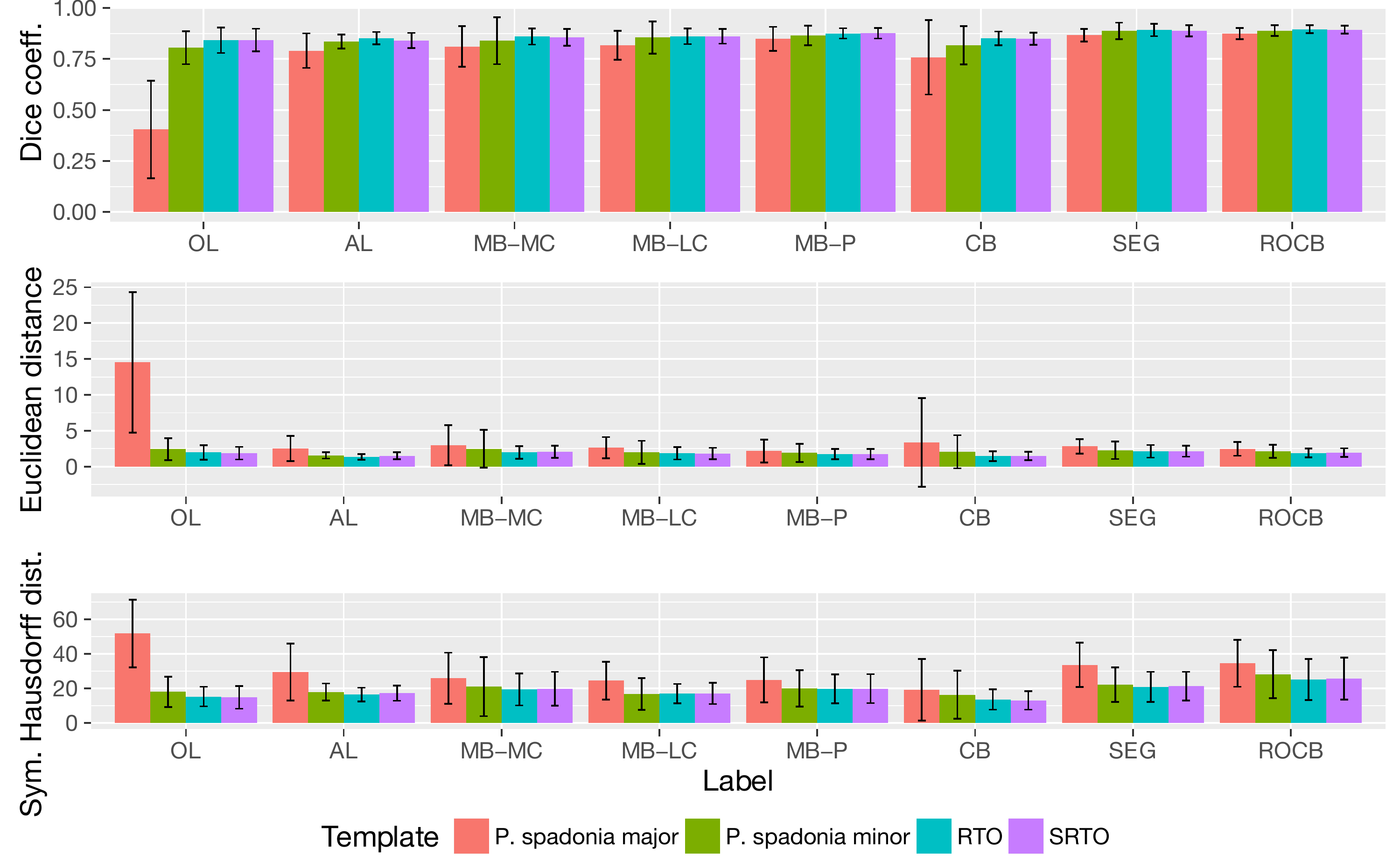}}
\caption{Evaluation of template performance per label. From top to bottom: Dice coefficient, Euclidean distance and Symmetric Hausdorff distance. Distances are expressed in microns.}
\label{fig:template-results}
\end{figure}

After analyzing the morphological differences of the sample populations based on their anatomical labels using the open-source toolbox MorphoLibJ \cite{legland2016morpholibj} (see Fig. \ref{fig:morphology-measures}), we decided to build and evaluate hybrid templates mixing minor samples of the different species. More specifically, we constructed one template (RTO) using all minor species except \textit{P. spadonia} (with 3 brains per species) and another template (SRTO) with \textit{P. spadonia} samples as well. All samples not used as part of templates, were used for testing their performance. Fig. \ref{fig:template-results} shows the evaluation results per label for the $4$ templates we created (\textit{P. spadonia} major, \textit{P. spadonia} minor, RTO and SRTO). The only template built with major samples performs notably worse than the other $3$ using both overlap and distance metrics, specially in the OL and CB. It is remarkable how the \textit{P. spadonia} minor performs only slightly worse than RTO and SRTO even though not a single \textit{P. spadonia} sample was used for testing.

\subsection{Evaluating worker polymorphisms and brain structure}
\begin{figure}[htb]
\centering
\centerline{\includegraphics[width=12cm]{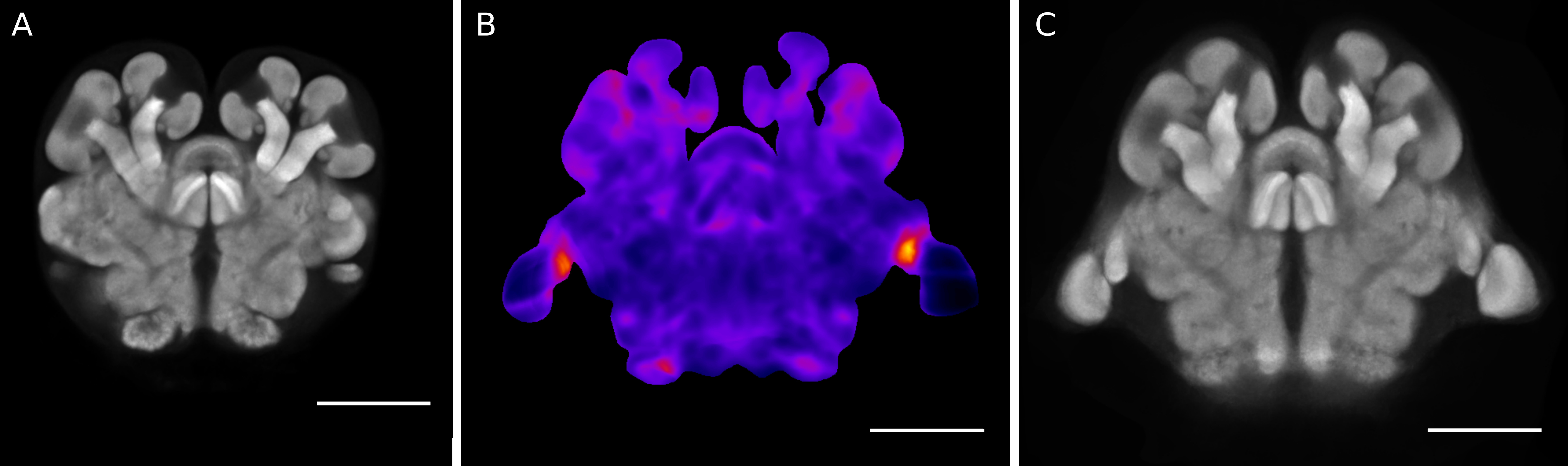}}
\caption{Inter-type deformation-based morphometry. From left to right: central view of \textit{P. spadonia} minor template, Jacobian determinant of deformation from minor to major template, and \textit{P. spadonia} major template. Scale bar: $100\mu m$.}
\label{fig:jacobian}
\end{figure}

One advantage of having templates of a single type of brain is that they allow to study the main morphological differences between species and/or subcastes. Following a methodology previously contrasted for fly brains \cite{manton2014combining}, we can register for instance our \textit{P. spadonia} minor and major templates to each other, and calculate the volume change of each voxel via the use of the Jacobian determinant. Once the difference in size is compensated with the affine transform, the local non-linear deformations can be visualized as a heatmap (see Fig. \ref{fig:jacobian}), emphasizing the regions of large differences.

\section{Conclusions and Future work}
We present a groupwise 3D registration strategy to build bias-free antbrain atlases that enable the efficient quantification of inter- and intraspecific variation in brain organization as evident in compartmental substructuring by automatic segmentation. We numerically evaluated template performance using expert-made manual annotations to validate that the atlases can be used to accurately study brain anatomy. To the best of our knowledge, this is the first time that automated atlases have been used to quantify ant brain volumes. The application of the current work to address questions in evolutionary neurobiology that require extensive datasets that adequately sample species-rich taxa will expedite the study of ant brain structure in relation to their ecological and evolutionary success and its association with division of labor and collective organization.

The ability to accurately and rapidly collect volumetric neuroanatomical data will greatly expand our ability to test social brain evolution in diverse clades such as ants.  Combined with phylogenetic analysis, immunohistochemistry, respirometry, high-performance liquid chromatography and other techniques, brain templates can help elucidate macroevolutionary and microevolutionary patterns of brain evolution, as well as mechanistic studies of the energetic cost of functionally specialized regions in the brain and the nature of aminergic control systems. This will allow to better understand regional brain investment in regard to the behavioral ecology of individual workers and their task specializations, and the impact of social processes operating at the colony-level.

\section*{Acknowledgements}
SA was supported by a Marie Skłodowska-Curie Individual Fellowship
(BrainiAnts-660976). DGG, AHP and JFAT, NSF were supported by grant IOS 1354291.

\bibliographystyle{IEEEbib}
\bibliography{refs}

\end{document}